\documentclass[
reprint,
amsmath,amssymb,
aps,
prb,
]{revtex4-2}

\usepackage{blindtext}
\usepackage[dvipsnames]{xcolor}

\usepackage{graphicx}
\usepackage{dcolumn}
\usepackage{bm}
\usepackage{hyperref}
\hypersetup{
colorlinks = true,
linkcolor = RoyalBlue,
citecolor = RoyalBlue,
urlcolor = Maroon,
}

\newcommand{\vecr}{\mathbf{r}}
\newcommand{\vecc}{\mathbf{c}}
\newcommand{\vecd}{\mathbf{d}}
\newcommand{\vecp}{\mathbf{p}}
\newcommand{\vecR}{\mathbf{R}}

\newcommand{\vecG}{\mathbf{G}}

\begin{document}

\title{Transferable empirical pseudopotentials from machine learning}

\author{Rokyeon Kim}
\email{Email: rrykim@gmail.com}
\affiliation{Korea Institute for Advanced Study, Seoul 02455, Korea}

\author{Young-Woo Son}
\email{Email: hand@kias.re.kr}
\affiliation{Korea Institute for Advanced Study, Seoul 02455, Korea}

\date{\today}

\begin{abstract}
Machine learning is used to generate empirical pseudopotentials that characterize the local screened interactions in the Kohn-Sham Hamiltonian.
Our approach incorporates momentum-range-separated rotation-covariant descriptors to capture crystal symmetries as well as crucial directional information of bonds, thus realizing accurate descriptions of anisotropic solids.
Trained empirical potentials are shown to be versatile and transferable such that the calculated energy bands and wave functions without cumbersome self-consistency reproduce conventional {\it ab initio} results even for semiconductors with defects, thus fostering faster and faithful data-driven materials researches.
\end{abstract}

\maketitle

\section{Introduction}

First-principles calculations based on the density functional theory (DFT)~\cite{Hohenberg1964, Kohn1965} have become standard tools for studying the physical properties of materials~\cite{Jones1989, Jones2015a,Louie2021NatMat,Marzari2021NatMat}.
Recently, applications of machine learning (ML) techniques to various computational methodologies based on the DFT has brought forth a new set of tools for investigating materials at the quantum scale~\cite{LeCun2015,Jordan2015,Carleo2019RMP,Giustino2020JPM,Kulik2022ES}.
Such novel approaches have given rise to a rapidly growing field, offering new insights and significant potential for prediction and analysis of materials properties~\cite{Behler2007, Bartok2010, Thompson2015, Li2015, Botu2015, Artrith2016, Khorshidi2016, Chmiela2017, Zhang2018b, Wang2018, Ryczko2019, Chen2020, Nagai2020NCM, Dick2020NatComm, Kirkpatrick2021Science, Kang2022NCM}.

One of popular applications in those developments has been the accelerated computations of physical quantities such as total energies and atomic forces~\cite{Behler2007,Bartok2010, Thompson2015, Botu2015, Artrith2016, Khorshidi2016, Chmiela2017, Zhang2018b, Wang2018, Kang2022NCM}. 
By circumventing a part of various computationally demanding processes involved in DFT calculations, the ML techniques provide efficient ways for improving various simulation methods such as molecular dynamics.
On the other hand, 
the integration of ML has also brought improvements to the exchange-correlation functionals within DFT~\cite{Snyder2012,Li2015,Nagai2020NCM,Ryczko2019, Chen2020,Dick2020NatComm,Kirkpatrick2021Science}, which are central to describing the many-electron effects in quantum systems.

Despite these strides, applications of ML to obtain precise quantum mechanical electronic structures for entire phase space of interests remain relatively unexplored~\cite{Hegde2017, Chandrasekaran2019, Schutt2019, Tsubaki2020, Unke2021,Li2022, Gong2023}.
Previous studies have utilized ML to study the electronic structures of one-dimensional~\cite{Hegde2017}, slab~\cite{Chandrasekaran2019}, and molecular~\cite{Schutt2019, Tsubaki2020, Unke2021} systems. We also note that the neural network was used to generate better transferable local pseudopotentials~\cite{Kim2022PCCP}.
Only recently, a general ML framework to construct DFT Hamiltonian in the tight-binding approach has been developed~\cite{Li2022, Gong2023}.
Accurate quantum properties of solids such as energy bands and wave functions are central to design and discovery of new materials with desired properties. 
However, the traditional DFT methods require a large amount of resources, partly because of the unavoidable self-consistent condition, posing significant challenges in data-intensive materials researches.
Hence, there is a pressing need for a faster method that utilizes ML to accelerate electronic structure calculations without sacrificing the accuracy of first-principles methods.

Before the advent of {\it ab initio} methods based on the DFT, empirical pseudopotential method (EPM)~\cite{Phillips1958} has been widely used as a fast and efficient method to calculate the electronic structure of materials because of its formal simplicity as well as less demanding computational resources.
Despite its extensive use for various solids~\cite{Phillips1958,Brust1962,Brust1962a,Cohen1965,Cohen1966,Chelikowsky1976,Cohen1989}, the EPM has limitations such as inaccurate wave functions~\cite{Yang1974} and transferability issues of the obtained pseudopotential~\cite{Cohen1966, Chelikowsky1976, Yeh1994, Mader1994, Zunger1996}.
To improve EPM, Wang and Zunger proposed the local-density-derived EPM, which generates pseudopotentials by inverting the Kohn-Sham (KS) potential in DFT calculations~\cite{Wang1995}.
However, the potentials obtained from this approach still suffer from a lack transferability to a wide range of materials, and the use of the spherical approximation results in errors in band structures, particularly for anisotropic crystal structures~\cite{Fu1997a}.

In this paper, we propose a neural network model to generate universal empirical pseudopotentials (EPs) encompassing the local screened interactions in solids.
We demonstrate that our EPs are versatile and transferable, reproducing conventional first-principles results for energy bands and wave functions without self-consistency.
To achieve this, we introduce a new set of rotation-covariant descriptors that capture the atomic and structural characteristics of target solids, enabling the transferability of learned EPs even to defective solids.
Moreover, our model seamlessly integrates into existing computational packages and can be extended to use advanced {\it ab initio} methods to calculate optical and transport properties.
%Thanks to our strategy for momentum separated descriptors, on top of bypassing the self-consistency, an additional reduction of resources is also demonstrated. 
We also demonstrate an extension to incorporate the non-local correlation effects without increasing computational complexity and resources.
Overall, our method accurately reproduces converged KS Hamiltonians without self-consistency, thus providing a reliable platform for data-intensive materials researches. 

\section{Machine learning framework}

In DFT using the local approximation for exchange-correlation functionals, the electronic structure of a solid is obtained by solving the KS equation~\cite{Kohn1965},
\begin{equation}
\left[ -\frac{1}{2} \nabla^2 + V_\mathrm{PS,nloc} + V(\vecr) \right] \psi_i = \epsilon_i \psi_i,
\end{equation}
where $\psi_i$ is the $i$-th KS wave function with an energy eigenvalue of $\epsilon_i$, 
$V_\mathrm{PS,nloc}$ the nonlocal part of the pseudopotential,
and $V(\vecr)$ the local potential for all interactions experienced by a single electron.
Specifically, $V(\vecr)$ is the sum of three terms: the local pseudopotential, Hartree potential, and exchange-correlation potential. In the DFT approach, $V(\vecr)$ for a given solid should be determined self-consistently.
Instead of the conventional way of computing converged potentials, 
we are to employ ML techniques to generate atomic EPs of $v(\vecr)$ for individual atoms, such that their summation equals the local potential of $V(\vecr)$ in a crystal. (we use $v$ and $V$ to denote atomic and crystalline potentials, respectively).
If the learned EPs can closely approximate the KS potentials, 
the KS Hamiltonian can be constructed only once, eliminating the self-consistency condition.

Specifically, let us consider a crystal structure denoted by $\mathcal C$, whose self-consistent local potential is $V_{\mathcal C}(\vecr)$.
We aim to find EPs of $v_{\mathcal C}^{\alpha}(\vecr)$ of the $\alpha$-th atom in the crystal structure $\mathcal C$ that satisfies
\begin{equation} \label{eq}
V_{\mathcal C}(\vecr) = \sum_{\vecR} \sum_\alpha v_{\mathcal C}^{\alpha}(\vecr-\vecR-\boldsymbol{\tau}_\alpha),
\end{equation}
where $\vecR$ is the Bravais lattice vector of the crystal, and $\boldsymbol{\tau}_\alpha$ is the position of the $\alpha$-th atom in the unit cell.
%We note that $\alpha$ is the index for each atom in the unit cell, not for atomic species, such as Si and O.
In the momentum space, Eq.~(\ref{eq}) can be written as
\begin{equation}
\label{eq3}
V_{\mathcal C}(\vecG_{\mathcal C})= \sum_{\alpha} S_{\mathcal C}^{\alpha}(\vecG_{\mathcal C}) v_{\mathcal C}^{\alpha} (\vecG_{\mathcal C}),
\end{equation}
where 
$S_{\mathcal C}^{\alpha}(\vecG_{\mathcal C})=\frac{1}{\Omega_{\mathcal C}} e^{-i \vecG_{\mathcal C} \cdot \boldsymbol{\tau}_\alpha}$ is the structure factor of the $\alpha$-th atom, ${\Omega_{\mathcal C}}$ the unit cell volume of structure $\mathcal C$, and
$\vecG_{\mathcal C}$ the reciprocal lattice vector of structure $\mathcal C$.

%\begin{equation}
%v_{\mathcal C}^{\alpha}(\vecG_{\mathcal C}) = \int d\vecr \, v_{\mathcal C}^{\alpha}(\vecr) e^{-i \vecG_{\mathcal C} \cdot \vecr}
%\end{equation}

For varying crystal structures, the EP of $v_{\mathcal C}^{\alpha}(\vecG_{\mathcal C})$ should depend on the local environment around the $\alpha$-th atom because the local symmetry, atomic coordination, and bond characteristics for each atom alter substantially.
To account for these factors, we introduce a descriptor of $\vecd_{\mathcal C}^{\alpha}$, whose specific form will be discussed later.
By incorporating this descriptor into the EP, we express the potential as functions of $\vecd_{\mathcal C}^{\alpha}$ and $\vecG_{\mathcal C}$, respectively.
So, resulting EPs can be written as $v[\vecd_{\mathcal C}^{\alpha}] (\vecG_{\mathcal C})$.
This is the universal EP, which must be transferable across different systems due to its ability to capture the complex local environments through the descriptor of $\vecd_{\mathcal C}^{\alpha}$.

\begin{figure}
\includegraphics[width=\columnwidth]{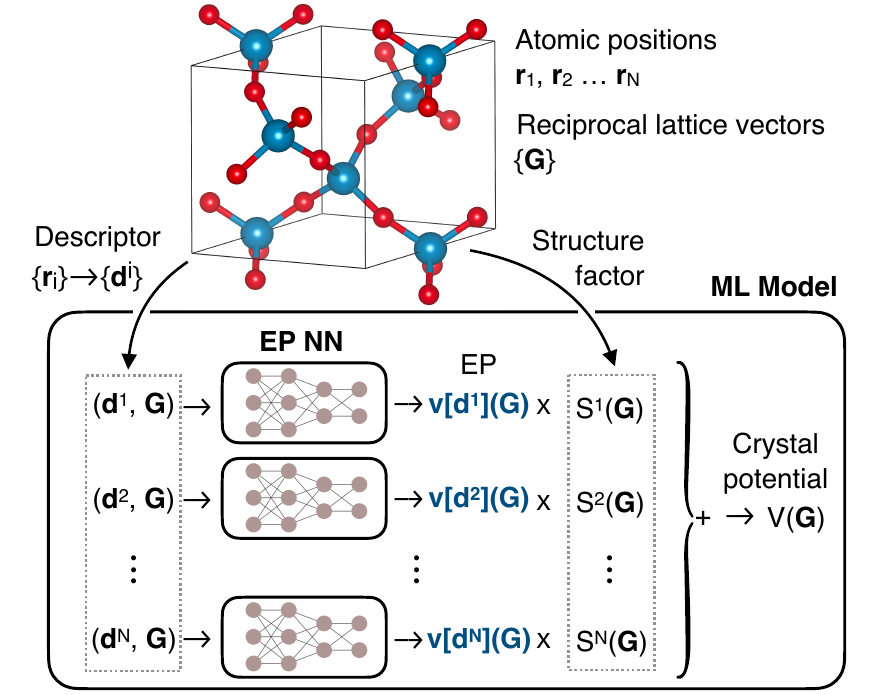}%
\caption{
The structure of a ML model
for EPs. From a crystal structure having $N$ atoms, we calculate $N$ descriptors of $\vecd^{i}$ representing the local environment at each atom. 
These descriptors, along with momentum $\vecG$, are fed into the EP neural network (EP NN) as input. 
The resulting $v[\vecd^{i}](\vecG)$ are multiplied by the structure factors of $S^{i}(\vecG)$, then summed over all atoms to yield the crystal potential of $V(\vecG)$.
}
\label{fig1_}
\end{figure}

With the introduction of the descriptor, we can apply ML to generate the EPs.
Since the KS potentials from DFT calculations are to be learned, we employ a ML model, as depicted in Fig.~\ref{fig1_}, similar to the high-dimensional neural network introduced by Behler and Parrinello \cite{Behler2007}.
Our model employs a weight-sharing neural network (denoted as EP NN in Fig.~\ref{fig1_}) to represent the EP of $v[\vecd](\vecG)$ with inputs of $\vecd$ and $\vecG$  (dropping the indices $\alpha$ and $\mathcal C$ for simplicity).
We optimize the ML model by minimizing the loss function for various crystal structures $\mathcal C$.
The loss function compares the DFT crystalline potential of $V_{\mathcal C, \mathrm{DFT}}(\vecG_{\mathcal C})$  with the corresponding ML-predicted $V_{\mathcal C, \mathrm{ML}}(\vecG_{\mathcal C})$ for a trained structure $\mathcal C$:
\begin{equation}
\mathrm{Loss} = \frac{1}{N_D} \sum_{\{\mathcal C\}}
\sum_{\vecG_{\mathcal C}} |V_{\mathcal C, \mathrm{DFT}}(\vecG_{\mathcal C})-V_{\mathcal C, \mathrm{ML}}(\vecG_{\mathcal C})|^2,
\end{equation}
where $N_D$ is the number of the training data and 
$\{{\mathcal C}\}$ indicates the summation for all training structures. 
Here, $V_{\mathcal C, \mathrm{ML}}(\vecG_{\mathcal C})$ is calculated by
\begin{equation}
\label{mleq}
V_{\mathcal C, \mathrm{ML}}(\vecG_{\mathcal C})= \sum_{\alpha} S_{\mathcal C}^{\alpha}(\vecG_{\mathcal C}) v[\vecd_{\mathcal C}^{\alpha}] (\vecG_{\mathcal C}),
\end{equation}
where $S_{\mathcal C}^{\alpha}(\vecG_{\mathcal C})$ can be readily calculated using the reciprocal lattice vectors and atomic positions.
After the ML procedure, we have the EP of $v[\vecd](\vecG)$ for general $\vecd$ and $\vecG$, which allow us to predict band structures for various new crystal structures.

To construct $V_{{\mathcal C},\mathrm{ML}}$ reflecting the crystal symmetry of a given structure $\mathcal C$, we introduced a momentum-dependent rotation-covariant descriptor that is modified from the atom-density representation~\cite{Bartok2013}.
In the representation~\cite{Bartok2013}, 
the sum of Gaussian functions with a variance of $\sigma^2$ is assigned as a density of $\rho^Z$ for each atomic species $Z$ inside a cutoff radius of $r_\mathrm{cut}$ such as 
$
\rho^{Z}(\vecr) = \sum_{\alpha \in Z} \exp\left[ -\frac{(\vecr-\vecr_\alpha)^2}{2 \sigma^2} \right].
$
Here the origin sets to the point where the local environment is evaluated.
With this density, we use two different types of descriptors as follows.

The first type is the density coefficient (DC) descriptor, denoted as $\vecc$, of which an element of $c_{lmn}^{Z}$ is defined as
%\begin{equation}
$
c_{lmn}^{Z} = \int d\vecr\, g_{n}^{*}(r) Y_{lm}^{*}(\hat{\vecr}) \rho^{Z}(\vecr)
$
%\end{equation}
where $g_n(r)$ is a set of radial basis functions, and $Y_{lm}(\hat{\vecr})$ spherical harmonics.
For practical applications, we limit the expansion using suitable $n_\mathrm{max}$ and $l_\mathrm{max}$ values.
The second type is the smooth overlap of atomic positions (SOAP) descriptor~\cite{Bartok2013}, denoted as $\vecp$, whose elements are defined as 
%\begin{equation}
$
p_{nn'l}^{ZZ'} = \pi \sqrt{\frac{8}{2l+1}} \sum_{m} \left(c_{nlm}^{Z}\right)^{*} c_{n'lm}^{Z'}.
$
%\end{equation}
With these definitions, the descriptor $\vecd$ can be either $\vecc$ or $\vecp$ depending on the purpose of ML
%$\vecG_{\mathcal C}$ 
as discussed below.
We note that only the SOAP descriptor seems to be widely used
% although it can be derived from the DC descriptor
 because its rotation-invariant nature may be suitable to represent scalar quantities of solids such as energy.
 However, it may not be adequate for describing the potential because it cannot reflect directional information.
We also note that, while the descriptor $\vecp$ was originally designed for the kernel methods \cite{Bartok2013}, it can be used as an input for a neural network \cite{Kocer2019}.

We use the DC descriptor for the EPs because, under a general rotation operation of $\mathcal{R}$, $v[\vecd](\vecG)$ should satisfy $v[\vecd](\vecG) = v[\mathcal{R} \vecd](\mathcal{R} \vecG)$, i.e., 
both $\vecd$ and $\vecG$ must be rotated simultaneously to yield the same $v$.
The DC descriptor $\vecc_{\mathcal C}^{\alpha}$ is rotation-covariant such that it is rotated by the well-defined rules using the Wigner $D$-matrices \cite{Bartok2013}. 
Therefore, 
we can use $\vecc_{\mathcal C}^{\alpha}$ as a main descriptor for learning EPs that can contain all local information for a crystal structure $\mathcal C$. 
On the other hand, with the SOAP descriptor $\vecp_{\mathcal C}^{\alpha}$, 
the EP is rotation-invariant, only depending on a magnitude ($G_{\mathcal C}$) of $\vecG_{\mathcal C}$ such that the resulting EP can be written as $v[\vecp_{\mathcal C}^{\alpha}](G_{\mathcal C})$.
In this case, instead of Eq.~(\ref{mleq}), we use the ML potential written as,
\begin{equation}
\bar{V}_{\mathcal C, \mathrm{ML}}(G_{\mathcal C})=\sum_{\alpha} \bar{S}_{\mathcal C}^{\alpha}(G_{\mathcal C}) v[\vecp_{\mathcal C}^{\alpha}](G_{\mathcal C}),
\end{equation}
where $\bar{S}_{\mathcal C}^{\alpha}(G_{\mathcal C})$ is the spherically averaged quantity of $S_{\mathcal C}^{\alpha}(\vecG_{\mathcal C})$.
Then, in the loss function, we compare $\bar{V}_{\mathcal C, \mathrm{ML}}(G_{\mathcal C})$ with the spherical averaged  $\bar{V}_{\mathcal C, \mathrm{DFT}}(G_{\mathcal C})$.

\iffalse
We introduce spherically-averaged quantities $\bar{V}_{\mathcal C}(G_{\mathcal C})$ and $\bar{S}_{\mathcal C}^{\alpha}(G_{\mathcal C})$ derived from $V_{\mathcal C}(\vecG_{\mathcal C})$ and $S_{\mathcal C}^{\alpha}(\vecG_{\mathcal C})$, respectively, as follows:
\begin{equation}
\bar{V}_{\mathcal C}(G_{\mathcal C})=\frac{1}{N_{G_{\mathcal C}}} \sum_{|\vecG_{\mathcal C}|=G_{\mathcal C}} V_{\mathcal C}(\vecG_{\mathcal C})
\end{equation}
\begin{equation}
\bar{S}_{\mathcal C}^{\alpha}(G_{\mathcal C})=\frac{1}{N_{G_{\mathcal C}}} \sum_{|\vecG_{\mathcal C}|=G_{\mathcal C}} S_{\mathcal C}^{\alpha} (\vecG_{\mathcal C}),
\end{equation}
where ${N_{G_{\mathcal C}}}$ counts the number of $\vecG_{\mathcal C}$ such that $|\vecG_{\mathcal C}|=G_{\mathcal C}$.
Due to the property $V_{\mathcal C}(-\vecG_{\mathcal C}) = V^{*}_{\mathcal C}(\vecG_{\mathcal C})$, $\bar{V}_{\mathcal C}(G_{\mathcal C})$ is real-valued. 
The same property holds true for $\bar{S}_{\mathcal C}(G_{\mathcal C})$ as well.
\fi

\begin{figure}
\includegraphics[width=\columnwidth]{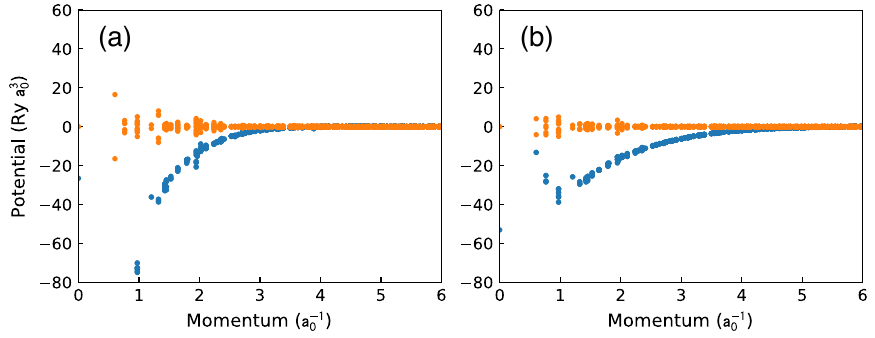}%
\caption{
Shape of EPs for (a) Si and (b) O in $\alpha$-quartz SiO$_2$. 
Blue (orange) dots represent real (imaginary) parts of the potentials.
Here, we directly inverted Eq.~(\ref{eq3}) without a ML procedure to obtain the potentials~\cite{note}.
The vertical distribution of data points indicates multiple momentum vectors for each momentum value, showing the directional dependence of the potentials.
For large momentum values, real values align along a line and imaginary values are suppressed, which can be also observed in other SiO$_2$ structures.
}
\label{fig2_}
\end{figure}

To ensure both efficiency and accuracy of our ML model, we adopt a hybrid approach using two distinct ML models that treats the directional dependency of the EP differently depending on the momentum values. 
For large $G$, the detailed directional information is found to be less relevant so that the spherical symmetric form suffices to construct EPs (see Fig~\ref{fig2_}). 
Therefore, the descriptor $\vecp$ is adopted to represent the potential for large $G$.
For small $G$, however, the directional information becomes crucial so that the descriptor $\vecc$ is used. 
Formally, in the hybrid ML process, the form $v[\vecp](G)$ is used for $G > G_\mathrm{cut}$, while $v[\vecc](\vecG)$ for $G \leq G_\mathrm{cut}$, where $G_\mathrm{cut} = 3 \, a_0^{-1}$ ($a_0$ is the Bohr radius).

\section{Applications to SiO$_2$}

Having established the universal EPM, we demonstrate a few examples showing the versatility of our ML model.
We choose silicon dioxide (SiO$_2$) or silica as an example owing to its diverse crystal structures and wide-ranging applications~\cite{Heaney1994}. 
To generate data for the ML model, we prepared 22 stable polymorphs of SiO$_2$ from the Materials Project \cite{Jain2013} as listed in Table~\ref{tab:str}.
\begin{table*}
\caption{
Unit cells of SiO$_2$ used for the generation of machine learning data.
}
\label{tab:str}
\begin{ruledtabular}
\begin{tabular}{ccc|ccc}
	Materials Project ID         & Number of atoms & Space group &
	Materials Project ID         & Number of atoms & Space group \\ \hline
    mp-6922    &  9  &      $P6_{2}22$     (No. 180)  &
	mp-10851   &  9  &      $P6_{4}22$     (No. 181)  \\
    mp-6930    &  9  &      $P3_{2}21$     (No. 154)  &
	mp-10948   & 12  &        $Pbcn$       (No. 60) \\  
	mp-6945    & 12  &    $P4_{1}2_{1}2$   (No. 92)  &
	mp-546794  & 12  &     $I\bar{4}2d$    (No. 122) \\
    mp-6947    &  6  &     $P4_{2}/mnm$    (No. 136) &
	mp-555235  & 24  &         $Cc$        (No. 9)  \\
	mp-7000    &  9  &      $P3_{1}21$     (No. 152)  &
	mp-555891  & 12  &       $P2_{1}$      (No. 4)  \\
    mp-7087    & 12  &     $P6_{3}/mmc$    (No. 194) &
	mp-556961  & 24  &  $P2_{1}2_{1}2_{1}$ (No. 19)  \\
	mp-7648    & 24  &      $C222_{1}$     (No. 20)  &
	mp-559091  & 12  &      $P6_{3}22$     (No. 182)  \\
    mp-7905    & 12  &        $Ibam$       (No. 72)  &
	mp-640917  &  9  &      $P3_{2}21$     (No. 154) \\ 
	mp-8059    & 24  &      $P2_{1}3$      (No. 198)  &
	mp-669426  & 24  &      $P2_{1}/c$     (No. 14)  \\
    mp-8352    & 24  &     $Fd\bar{3}m$    (No. 227) & 
	mp-972808  & 12  &       $P2_{1}$      (No. 4)  \\ 
	mp-9258    & 12  &     $Pa\bar{3}$     (No. 205)  &
	mp-1071820 & 12  &        $Ibam$       (No. 72)  \\
\end{tabular}
\end{ruledtabular}
\end{table*}
For each sample, we constructed $p \times q \times r$ supercells ($p$, $q$, and $r$ are integers) limiting the maximum number of atoms in the supercell to 24. 
We then randomly perturbed the length of the supercell lattice vectors and the internal coordinates of the atoms within 5\% to increase the diversity of the generated data. 
For 10\% of the learning samples, we create arbitrary defective structures by randomly removing atoms in the unit cell. 
In total, we prepared 10,224 inputs for DFT calculations. 
For each input $\mathcal C$, we calculated $V_{{\mathcal C}, \mathrm{DFT}}$, 
$S_{\mathcal C}^\alpha$, $\vecc_{\mathcal C}^\alpha$ and $\vecp^\alpha_{\mathcal C}$, which were used to train the ML model.
We implemented our universal EPMs in the {\sc Quantum Espresso} package~\cite{QE}, which we also used for performing reference DFT calculations.
Norm-conserving LDA pseudopotentials from the PseudoDojo \cite{vanSetten2018} were utilized.
The $k$-point mesh was set with approximately 1,000 $k$-points per reciprocal atom and the cut-off energy for the plane-wave basis set was 84 Ry.

We used an in-house modified version of the DScribe package \cite{Himanen2022} to calculate descriptors.
Gaussian functions with $\sigma=1$~{\AA} were employed for the density. 
For the basis functions, we used Gaussian type orbitals and real spherical harmonics as implemented in the package.
The DC descriptor $\vecc$ was calculated with $r_\mathrm{cut}=10$~{\AA}, $n_\mathrm{max}=7$, $l_\mathrm{max}=7$,
resulting in a total of 896 features.
For the SOAP descriptor $\vecp$, we used $r_\mathrm{cut}=10$~\AA, $n_\mathrm{max}=6$, $l_\mathrm{max}=6$,
and applied a weighting function of the form 
$c (1+2(\frac{r}{r_0})^3-3(\frac{r}{r_0})^2)^m$ with $c=1$, $m=2$, $r_0=10$ {\AA} \cite{Caro2019},
leading to a total of 546 features.

For the EP NN (see Fig.~\ref{fig1_}), we employed fully connected neural networks with the following structures.
For the input $(\vecc, \vecG)$ comprising the DC descriptor $\vecc$ and the vector $\vecG$, we employed 3 hidden layers, each consisting of 1024 neurons with rectified linear unit (ReLU) activation function. The output layer was composed of two neurons, representing the real and imaginary parts of the potential $v[\vecc](\vecG)$, respectively.
Similarly, for the input $(\vecp, G)$ comprising the SOAP descriptor $\vecp$ and the scalar $G$, we utilized 3 hidden layers, each consisting of 512 neurons with ReLU activation function. The output layer consisted of a single neuron representing the potential $v[\vecp](G)$. TensorFlow \cite{tensorflow2015-whitepaper} with the Keras API \cite{chollet2015keras} was utilized to implement the entire ML process.

\begin{figure}
\includegraphics[width=\columnwidth]{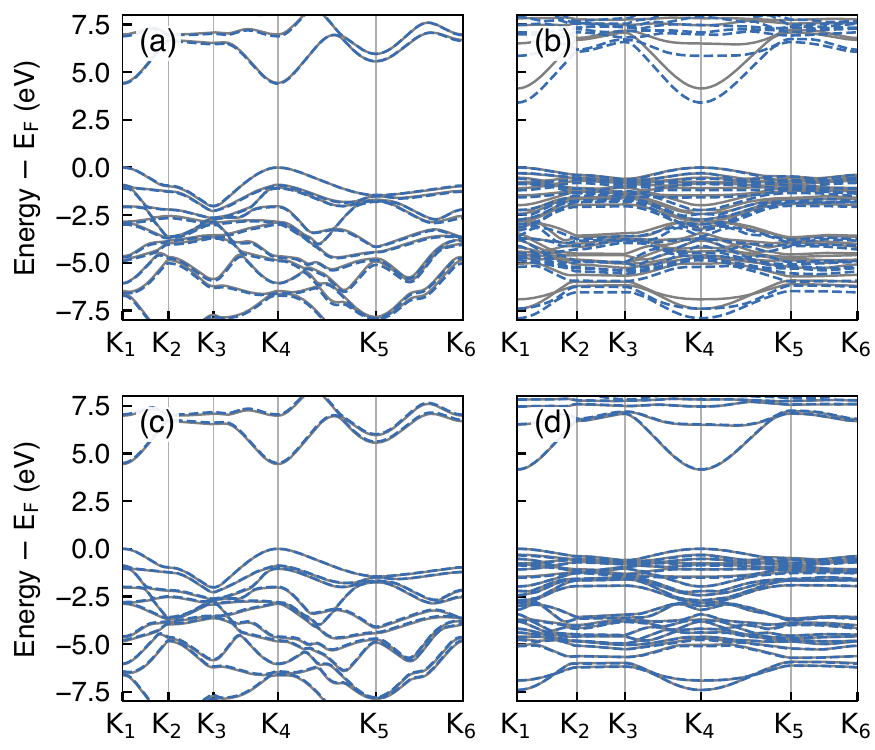}%
\caption{
Band structures of two representative SiO$_2$ structures, C1 and C2 (see text).
Gray solid and blue dashed lines represent the DFT and EPM results, respectively.
${\mathrm K}_i(i=1,\cdots,6)$ are reduced coordinates in reciprocal space, representing (0, 0, 0), (0.5, 0 ,0), (0.5, 0.5, 0), (0, 0, 0), (0.5, 0.5, 0.5), (0.5, 0, 0) in their corresponding crystal structures, respectively. $E_F$ denotes the Fermi energy.
The full-spherical approximation results for (a) C1 and (b) C2.
The hybrid model results for (c) C1 and (d) C2.
}
\label{fig3_}
\end{figure}

Figure~\ref{fig3_} shows the the band structures of two SiO$_2$ structures denoted as C1 and C2, which were created by randomly perturbing the unit cell vectors and internal coordinates of stishovite (space group $P4_2/mnn$) and fibrous (space group $Ibam$) polymorphs, respectively.
We compare the band structures obtained using the full-spherical approximation and the hybrid model. 
In the full-spherical approximation, we assumes the form $v[\vecp](G)$ for all momentum values as opposed to the hybrid model. Figures~\ref{fig3_}(a) and \ref{fig3_}(b) show the band structures using the full-spherical approximation for C1 and C2, respectively. 
For C1, the full-spherical approximation yields satisfactory agreement with its DFT band dispersions, but for C2, we observe discrepancies in the band structures because of its anisotropic crystal structure. 
Figures~\ref{fig3_}(c) and \ref{fig3_}(d) show the energy bands of C1 and C2, respectively, using the hybrid model. 
The hybrid model significantly improves the accuracy of the EPM predictions of bands, being identical to those from DFT. 
These results emphasize the importance of incorporating directional information in the EPs to accurately describe the band structure, especially for anisotropic materials.

\begin{figure*}
\includegraphics[width=2.05\columnwidth]{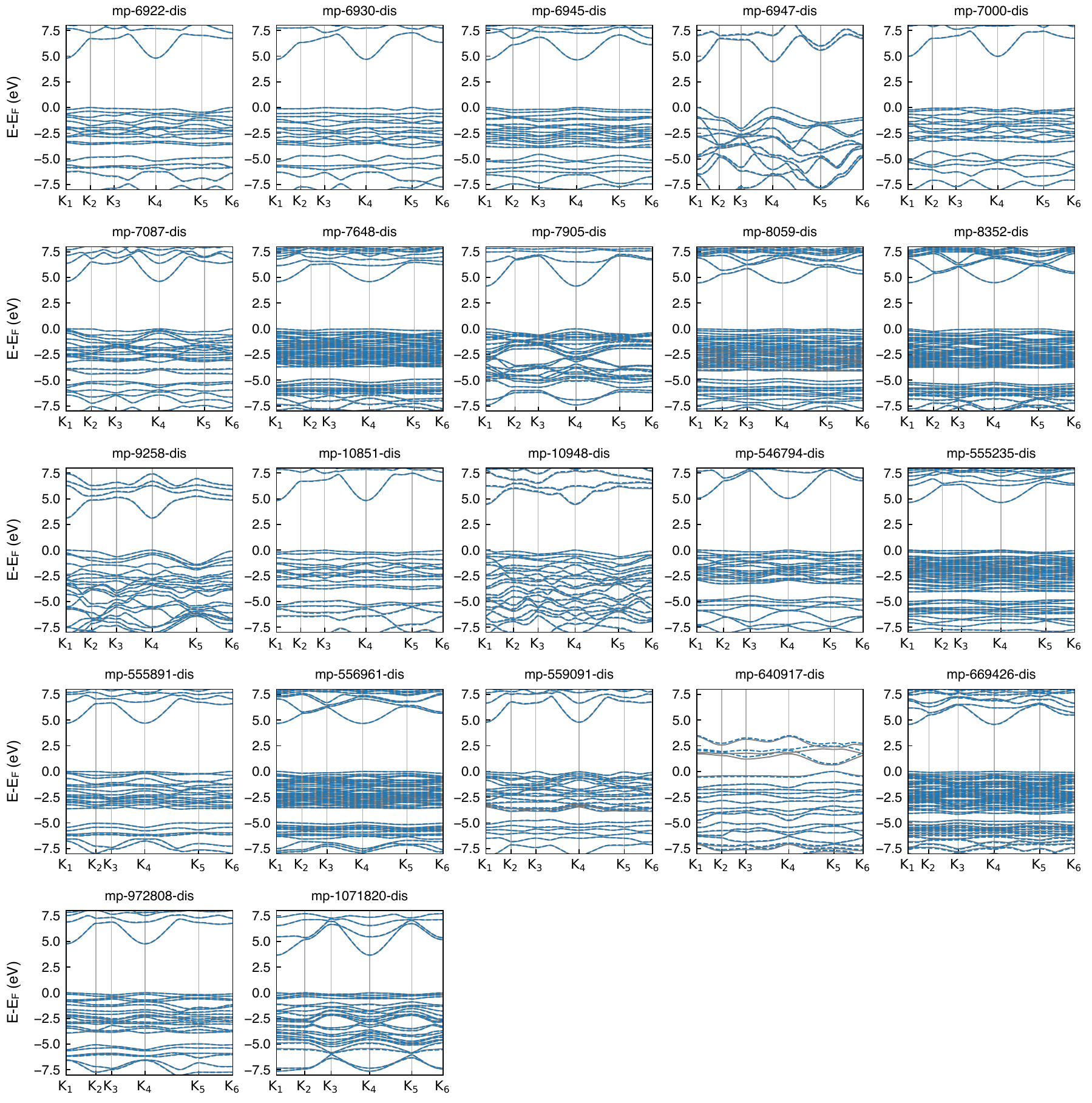}%
\caption{
Band structures for SiO$_2$ with Material Project IDs ranging from mp-6922 to mp-1071820 (see Table~\ref{tab:str}), 
where the length of the lattice vectors and internal coordinates for each structure are randomly disturbed within 5\% (denoted as ``-dis'' in labels).
${\mathrm K}_i$ $(i=1,\cdots,6)$ are reduced coordinates in reciprocal space, representing (0, 0, 0), (0.5, 0 ,0), (0.5, 0.5, 0), (0, 0, 0), (0.5, 0.5, 0.5), (0.5, 0, 0) in their corresponding crystal structures, respectively.
Solid gray and dashed blue lines represent the DFT and the EPM results, respectively.
}
\label{fig4_}
\end{figure*}

For the crystal structures listed in Table~\ref{tab:str}, the EPM band structures calculated from the hybrid model are displayed in Fig.~\ref{fig4_}.
Here, we adjusted each unit cell's lattice vectors and internal coordinates randomly perturbed by up to 5\% to test the transferability and predictive capability of our EPM.
In comparing the band structures, we observe a good agreement between the EPM and DFT results. 
An exception is noted for mp-640917 in the conduction bands, attributed to its highly anisotropic nature.

\begin{figure}
\includegraphics[width=\columnwidth]{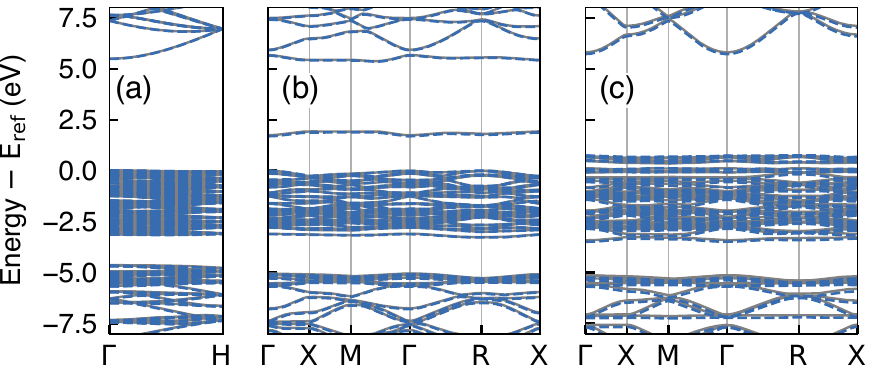}%
\caption{
(a) The band structure of a crystal with 72 atoms in a unit cell (see details in text).
The band structures of defective cells for (b) oxygen vacancy and (c) silicon vacancy.
Solid gray and dashed blue lines represent the DFT 
and the EPM results, respectively.
In (a), $E_{\mathrm{ref}}=E_F$.
In (b), we set $E_{\mathrm{ref}}$ at the band energy of the $(N-2)$-th electrons, while in (c) $(N+4)$-th electrons, where $N$ is the number of electrons in each crystal structure.
}
\label{fig5_}
\end{figure}

In Fig.~\ref{fig5_}(a), we present the band structure of a crystal having 72 atoms in its unit cell (Materials Project ID of mp-6930 with a space group of $P3_2 21$), which exhibits excellent agreement between the EPM and DFT results.
It is worth noting that our hybrid ML model has been trained on a data set containing no more than 24 atoms in unit cell, and yet it accurately predicts the energy bands of a crystal with the significantly larger cell. 
This finding highlights the potential of our method in reliably predicting the electronic properties of materials of larger cell size, even with trained EPs from smaller cell sizes.

As shown in all bands from EPM in Figs.~\ref{fig3_}, \ref{fig4_}, and \ref{fig5_}, we confirm that our hybrid EPM correctly predicts the band degeneracies at all high-symmetry points, indicating that the EP has learned the required crystalline symmetry for the potentials. 
Although our model does not include the symmetry components unlike equivariant neural network methods~\cite{Thomas2018, Anderson2019, Fuchs2020, Batzner2022, Gong2023}, we were able to learn the correct symmetry by designing a suitable rotation-covariant descriptor, generating data, and developing a ML model.

Figures~\ref{fig5_}(b) and \ref{fig5_}(c) present the energy bands of SiO$_2$ with oxygen and silicon vacancies, respectively. We confirm that the defective structures for our test run do not belong to the learning set.
Our EPM accurately predicts the band structures and defect energy levels if compared with the DFT results.
Our study highlights the capability of EPM to accurately predict the defect band structure with computational efficiency compared to DFT. 
%This result has significant implications for future studies of defect properties in larger structures.

\begin{figure}
\includegraphics[width=\columnwidth]{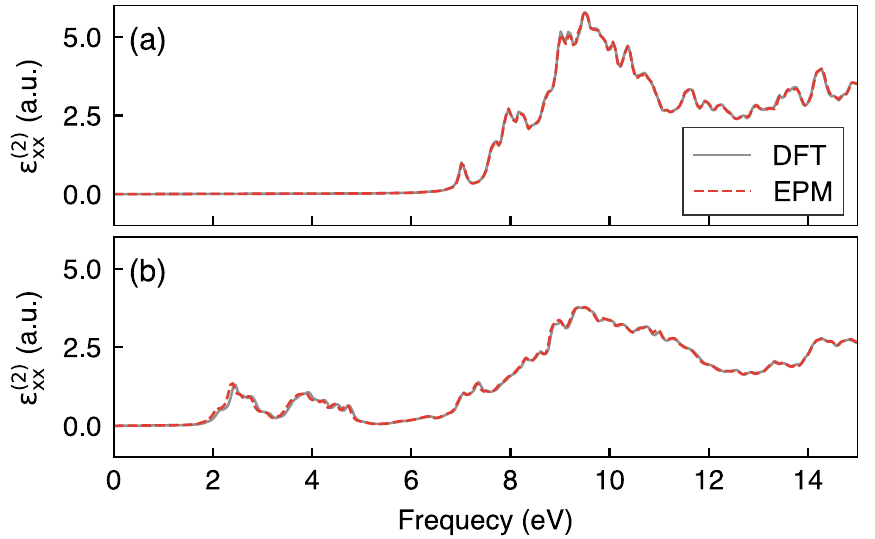}%
\caption{Frequency dependent non-interacting dielectric functions calculated for
(a) a pristine SiO$_2$ structure
(b) and the SiO$_2$ structure with an oxygen vacancy.
}
\label{fig6_}
\end{figure}

As a final example, we use both band energies and wave functions to compute the dielectric function.
Figures~\ref{fig6_}(a) and \ref{fig6_}(b) display the imaginary part of the non-interacting dielectric function of $\epsilon_{xx}^{(2)}(\omega)$ for a pristine SiO$_2$ and one with an oxygen vacancy, respectively, demonstrating the excellent agreement between matrix elements obtained from the DFT and EPM results.
%Importantly, our approach can accurately predict both the wavefunctions and band energies, showcasing its potential for predicting the optical properties of larger systems with complex crystal structures.

\section{Applications to silicon}

In this section, we detail calculations conducted on a few polymorphs of silicon. 
While we've crafted separate datasets and an ML model for Si, it's worth noting that they could be merged with those of SiO$_2$ to create a unified ML model applicable to both SiO$_2$ and Si. 
To generate data for the ML model, we prepared 3 stable polymorphs of Si as listed in Table~\ref{tab:Si}.
\begin{table}
\caption{
Unit cells of Si used for the generation of machine learning data.
}
\label{tab:Si}
\begin{ruledtabular}
\begin{tabular}{ccc}
Materials Project ID & Number of atoms & Space group \\
\hline
mp-149 & 8 & $Fd\bar{3}m$ (No. 227)  \\
mp-165 & 4 & $P6_3/mmc$ (No. 194) \\  
mp-1095269 & 24 & $Cmcm$ (No. 63) \\
\end{tabular}
\end{ruledtabular}
\end{table}
The unit cells were retrieved from the Materials Project \cite{Jain2013}.
For each sample, we constructed $p \times q \times r$ supercells ($p$, $q$, and $r$ are integers) limiting the maximum number of atoms in the supercell to 24. 
We then randomly perturbed the length of the supercell lattice vectors and the internal coordinates of the atoms within 5\% to increase the diversity of the generated data. 
In total, we prepared 2,496 inputs for DFT calculations. 

We employed the same parameters as those used in the SiO$_2$ to compute descriptors.
As a result of reducing the number of atomic species, $\vecc$ contains 448 features, and $\vecp$ contains 147 features.
For the EP NN, we utilized three hidden layers, each containing 512 neurons, for the input of $(\vecc, \vecG)$. Similarly, for the input of $(\vecp, G)$, we employed three hidden layers, each comprising 128 neurons.

\begin{figure}
\includegraphics[width=\columnwidth]{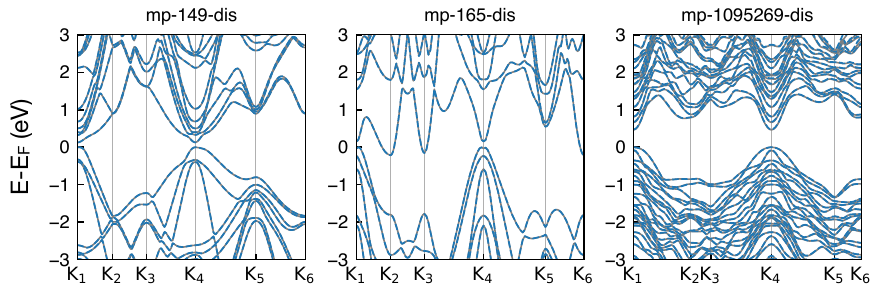}%
\caption{
Band structures for Si with Material Project IDs mp-149, mp-165, and mp-1095269 (see Table~\ref{tab:Si}), where the length of the lattice vectors and internal coordinates for each structure are randomly disturbed within 5\% (denoted as ``-dis'' in labels).
${\mathrm K}_i$ $(i=1,\cdots,6)$ are reduced coordinates in reciprocal space, representing (0, 0, 0), (0.5, 0 ,0), (0.5, 0.5, 0), (0, 0, 0), (0.5, 0.5, 0.5), (0.5, 0, 0) in their corresponding crystal structures, respectively.
Solid gray and dashed blue lines represent the DFT and the EPM results, respectively.
}
\label{fig7_}
\end{figure}

The band structures corresponding to the unit cells listed in Table~\ref{tab:Si} are illustrated in Fig.~\ref{fig7_}, where we used the hybrid model for the EPM. 
To assess our EPM, we introduced random variations to the lattice vectors and internal coordinates, with perturbations up to 5\%. 
For all three distinct structures, there's an excellent agreement between the DFT and EPM results, showing the reliability and accuracy of our approach.

\section{Discussion and Summary}

\begin{figure}
\includegraphics[width=\columnwidth]{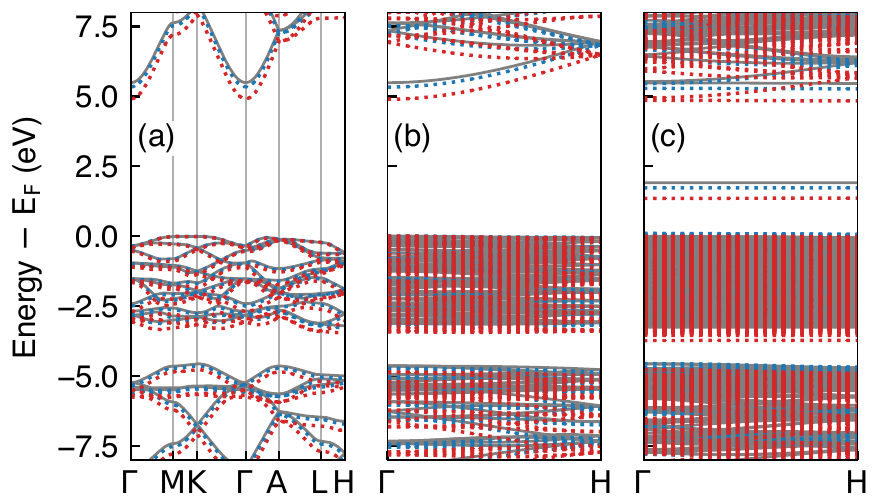}%
\caption{Band structures obtained with different energy cut-off values for (a) $\alpha$-quartz SiO$_2$ (9 atoms), (b) $2 \times 2 \times 2 $ supercell (72 atoms) and (c) $3 \times 3 \times 3 $ supercell with an oxygen vacancy (242 atoms).
For each figure, the solid gray lines represent the DFT results obtained using an energy cut-off of 84 Ry, while the dotted blue (red) lines correspond to the EPM results obtained using an energy cut-off of 50 (40) Ry.
}
\label{fig8_}
\end{figure}

Before concluding, we provide a couple of remarks on our universal EPM. Since the present ML model constructs a converged KS Hamiltonian in momentum space, a high energy cutoff is not necessary to converge density matrix. The only limiting factor for the cutoff comes from the kinetic energy. We demonstrate such an efficiency by reducing the cutoff for energy bands calculations of silica polymorphs. 
As shown in Fig.~\ref{fig8_}, in case of 40\% reduction of the cutoff for conventional DFT calculations, the almost unaltered bands can be obtained using only about 4\% of computational resources used for DFT.
Table~\ref{tab:time} compares the time needed for band structure calculations using DFT and EPM for three different systems.
For the unit cell of $\alpha$-quartz containing 9 atoms, the self-consistent DFT calculation using a $5 \times 5 \times 4$ $k$-mesh for electron density took 107 seconds, and the subsequent band structure calculation for 300 $k$-points required 226 seconds.
In contrast, at the same energy cut-off of 84 Ry, the EPM computed the band structure in 47 seconds.
Further reducing the energy cut-off to 50 Ry shortened this time to 26 seconds. 
When expanding to a larger 72-atom supercell using, the self-consistent DFT calculation for density using a $3 \times 3 \times 2$ $k$-mesh took 2,294 seconds.
The band structure calculation for 30 $k$-points were completed in 4,080 seconds.
However, using the EPM and energy cut-offs of 84 Ry and 50 Ry, the band structure computation times were reduced to 807 and 276 seconds, respectively.
\begin{table*}
\caption{Comparison of computation times for band structures.}
\label{tab:time}
\begin{ruledtabular}
\begin{tabular}{cccccccc}
Number of & Number of &  $k$-mesh  & Number of & \multicolumn{4}{c}{time (second)} \\ \cline{5-8}
atoms & CPU\footnote{Intel(R) Xeon(R) CPU E5-2690 v2 @ 2.800GHz (10 cores)} nodes &(density) & $k$-points (band) & DFT density & DFT band & EPM band (84 Ry) & EPM band (50 Ry) \\
\hline
9\footnote{$\alpha$-quartz SiO$_2$} 
& 1 & $5 \times 5 \times 4 $ & 300 & 107 & 226 & 47 & 26 \\
72\footnote{$2 \times 2\times 2$ supercell} 
& 2 & $3 \times 3 \times 2 $& 30 & 2,294 & 4,080 & 807 & 276\\  
242\footnote{$3 \times 3\times 3$ supercell with an oxygen vacancy} 
& 2 & $2 \times 2 \times 2 $ & 20 & 31,500 & 85,740 & 14,340 & 6,900 \\
\end{tabular}
\end{ruledtabular}
\end{table*}
We also note that, for a Hamiltonian matrix from our EPM, a desired set of energies and wave functions within a target energy window can be extracted~\cite{arpack1998, Polizzi2009}.
These merits would lead to a significant reduction of resources in making computation-driven 
database for specific materials properties.

\begin{figure}
\includegraphics[width=\columnwidth]{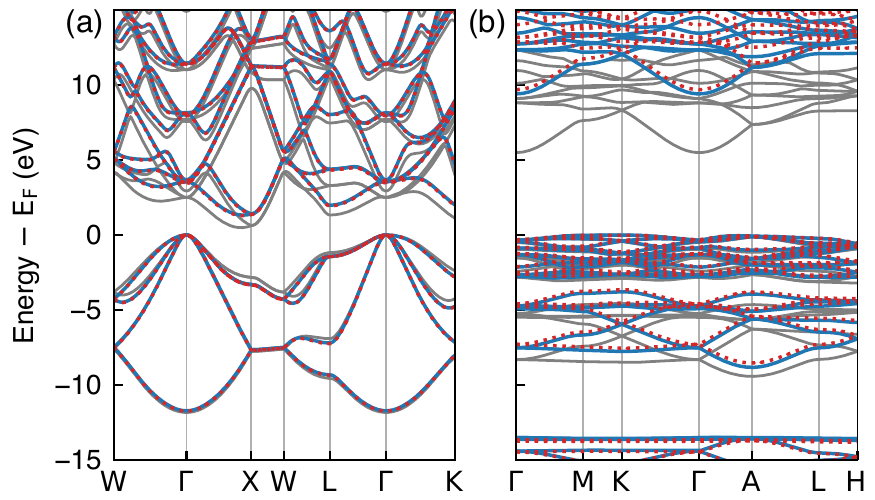}%
\caption{Band structures of (a) Si and (b) SiO$_2$ incorporating the effects of the on-site interaction $U$ and the inter-site interaction $V$.
In each figure, the solid gray lines represent LDA results, the solid blue lines denote self-consistent LDA+$U$+$V$ result, and the dotted red lines show the one-shot +$U$+$V$ results performed on top of the EPM results.
}
\label{fig9_}
\end{figure}

We also tested a possible extension to include non-local correlations within DFT.  
Recent studies~\cite{Campo2010,Lee2020,Tancogne2020,Yang2021}
have shown that the DFT with self-consistent inter-site Hubbard interactions can yield excellent many-body quasiparticle energy gaps  of semiconductors. 
Using the local charge density from our EPM, 
we performed one-shot calculations for the nonlocal interactions
of Si and SiO$_2$ and obtained their energy bands shown in Fig.~\ref{fig9_}, agreeing with results from the aforementioned methods very well.

We recognize certain limitations inherent in our framework. First, while our model demonstrates effectiveness in predicting electronic properties such as band structures, its capability to accurately predict total energies is not yet fully developed. The current accuracy of our ML model, approximately 99.9\%, does not meet the criteria required for reliable total energy predictions. Second, the training of our model has been focused on periodic systems of a single chemical composition, which restricts its direct inference to surfaces or systems with varied chemical compositions. Retraining with an expanded and more diverse dataset is required for the reliable predictions for these systems. The refinement of our model to improve total energy predictions and the development of a ML architecture and dataset to enhance extrapolation capabilities present opportunities for future research.

In summary, we present a new empirical computational method using neural network as accurate as {\it ab initio} methods based on DFT. Our method combines the merits of EPM with ML approaches by introducing a new hybrid descriptor reflecting local symmetries of materials reliably even when they have defects. 
We demonstrate the versatility and efficacy of our approach 
by computing energy bands of polymorphs of silica and silicon
as well as their defective structures.
Since transferable empirical pseudopotentials in our study can replace all local Hartree, atomic, and exchange-correlation potentials in KS Hamiltonians without self-consistency, results can be applied to all post-processing computational tools within existing first-principles calculations packages. Moreover, as demonstrated in defective SiO$_2$, our methodology holds promise for non-ideal structures. So, we also expect that our methods can be used to generate transferable empirical pseudopotential for molecules, clusters, and disordered systems.
We anticipate that our universal EPM will play important roles in constructing accurate computation-driven materials database.

\begin{acknowledgments}
Y.-W.S. was supported by the National Research Foundation of Korea (NRF) (Grant No. 2017R1A5A1014862, SRC program: vdWMRC center) and KIAS individual Grant No. (CG031509). Computations were supported by Center for Advanced Computation of KIAS.
\end{acknowledgments}

%\begin{widetext}
%\end{widetext}

% \begin{figure*}
% two column figure
% \end{figure*}

%\bibliography{epm}

\begin{thebibliography}{70}
\bibitem [{not()}]{note}%
  \BibitemOpen
  \href@noop {} {}\bibinfo {note} {For the system in which we aim to compute
  potentials, we generated $N_{\mathrm{str}}$ distinct crystal configurations
  by making minor adjustments to the internal coordinates. Subsequently, we
  formulated the matrix equation \begin{equation*} \begin{pmatrix} V^{}_{1}
  (\vecG) \\ : \\ V^{}_{N_\mathrm{str}} (\vecG) \end{pmatrix} = \begin{pmatrix}
  S^{\mathrm{Si}}_{1} (\vecG) & S^{\mathrm{O}}_{1} (\vecG) \\ : & : \\
  S^{\mathrm{Si}}_{N_\mathrm{str}} (\vecG) & S^{\mathrm{O}}_{N_\mathrm{str}}
  (\vecG) \end{pmatrix} \begin{pmatrix} v^{\mathrm{Si}} (\vecG) \\
  v^{\mathrm{O}} (\vecG) \end{pmatrix} \end{equation*} and inverted it to
  obtain potentials for Si and O.}\BibitemShut {Stop}%
\end{thebibliography}


\end{document}